\documentclass[]{spie}  

\usepackage{amsmath,amsfonts,amssymb}
\usepackage{graphicx}
\usepackage[colorlinks=true, allcolors=blue]{hyperref}
\usepackage{siunitx}

\newcommand{\orcid}[1]{\href{https://orcid.org/#1}{\includegraphics[width=10pt]{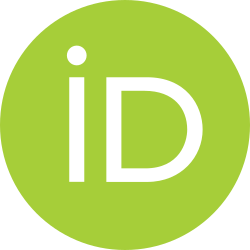}}}

\title{CCAT: LED mapping of the 850 GHz instrument module for Prime-Cam on FYST.}

\author[a,b]{James R. Burgoyne}
\author[a]{Jonah J. Lee}
\author[c]{Jordan D. Wheeler}
\author[c]{Jason E. Austermann}
\author[c]{James A. Beall}
\author[d,e]{Scott C. Chapman}
\author[a]{Anthony I. Huber}
\author[c]{Johannes Hubmayr}
\author[f]{Matthew Koc}
\author[c,f]{Joel Ullom}
\author[c,f]{Anna K. Vaskuri}
\author[c]{Michael R. Vissers}
\author[a]{Ruixan (Matt) Xie}
\author[]{the CCAT collaboration}

\affil[a]{Dept. of Physics and Astronomy, University of British Columbia, Vancouver, Canada}
\affil[b]{Dept. of Physics and Astronomy, University of Victoria, Victoria, Canada}
\affil[c]{National Institute of Standards and Technology, Quantum Sensors Division, Boulder, Colorado, United States}
\affil[d]{Dept. of Physics \& Atmospheric Science, Dalhousie University, Dalhousie, Canada}
\affil[e]{Herzberg Astronomy \& Astrophysics Research Centre, National Research Council Canada, Victoria, Canada}
\affil[f]{University of Colorado, Department of Physics, Boulder, Colorado, United States}
\authorinfo{Further author information: (Send correspondence to James R. Burgoyne)\\James R. Burgoyne: E-mail: jburgoyne@phas.ubc.ca}

\begin{document} 
\maketitle

\begin{abstract}
The 850 GHz instrument module for Prime-Cam on the Fred Young Submillimeter Telescope (FYST) utilizes large-format arrays of superconducting microwave kinetic inductance detectors (MKIDs) to enable high-density submillimeter surveys. However, fabrication-induced variations introduce resonant frequency scatter and spectral collisions, degrading readout fidelity and detector yield. This work presents the design, implementation, and characterization of a cryogenic, Charlieplexed LED mapping system developed as a fully engineered subsystem to systematically associate measured resonator frequencies with their physical positions on the array. Tested at 100 mK, the automated mapping methodology demonstrates high robustness, achieving a 99.7\% tracking yield by successfully mapping 11,576 out of 11,608 designed resonators across 11 active networks. The array exhibited a mean fractional frequency scatter of $3.43\pm0.92\%$, revealing macroscopic geometric gradients and fabrication artifacts across the focal plane. The mapping algorithm successfully resolved resonator-pixel coordinates, even under conditions of spectral overlap or localized LED failure, by leveraging optical spillover from adjacent pixels. Spatial variations in the coupling quality factor ($Q_c$) were also characterized, tracking an alternating-row pattern introduced by feedline routing asymmetries. These high-fidelity spatial-frequency maps provide the critical empirical baseline required for a post-fabrication lithographic capacitor trimming pipeline to eliminate frequency collisions and maximize final detector yield.
\end{abstract}


\section{INTRODUCTION}
\label{sec:intro}  

The CCAT Collaboration is developing the Fred Young Submillimeter Telescope (FYST), a \qty{6}{m} aperture facility optimized for observations at millimeter and submillimeter wavelengths from a high-altitude site \cite{parshley2018b}. One of its first-light instruments, Prime-Cam, will deploy multiple instrument modules spanning frequencies from \qty{280}{GHz} to \qty{850}{GHz}, enabling surveys targeting early-universe cosmology, galaxy evolution, and large-scale structure \cite{vavagiakis2018, aravena2023}. Each module employs large-format arrays of superconducting microwave kinetic inductance detectors (MKIDs), whose inherent multiplexing capability makes them well suited for scaling to the detector counts required for these science goals \cite{zmuidzinas2012}.

A central challenge in the operation of large KID arrays is the accurate placement of resonator frequencies within a limited readout bandwidth. While resonators are lithographically designed to occupy specific frequencies, fabrication-induced variations in material properties and geometry lead to significant scatter in the realized resonant frequencies. This scatter produces frequency collisions, in which two or more resonators overlap within some tolerance decided by science goals, usually a few linewidths, degrading readout fidelity and reducing the effective detector yield. As array sizes and multiplexing factors increase, the impact of such collisions becomes a dominant limitation on instrument performance.

LED mapping has emerged as a practical technique for associating measured resonator frequencies with their physical detector locations on the array \cite{middleton2025, martsen2025, albert2024, shroyer2023, liu2022}. By illuminating detectors sequentially with a spatially registered array of light-emitting diodes, the response of individual resonators can be correlated with known illumination positions, producing a mapping between frequency space and physical space. This mapping enables post-fabrication corrective processes, such as lithographic capacitor trimming, which adjust resonator frequencies to mitigate collisions and recover yield.

Previous implementations of LED mapping for KID arrays, including those developed for lower-frequency modules, have demonstrated the basic viability of this approach \cite{middleton2025}. However, as detector counts increase and tolerances tighten, particularly for higher-frequency modules such as the \qty{850}{GHz} array, the performance of the LED mapping system becomes increasingly sensitive to details of optical coupling, electrical addressing, timing synchronization, and data analysis methodology.

In this work, we present the design, implementation, and characterization of an LED mapping system developed for the MKID detector arrays in the \qty{850}{GHz} instrument module of Prime-Cam. This work treats the LED mapper as a fully engineered subsystem, with explicit design requirements derived from array geometry, cryogenic constraints, and mapping accuracy targets. We describe the printed circuit board (PCB) architecture, LED selection and layout, electrical control scheme, and mechanical integration with the detector package, along with measurements of optical uniformity and system-level performance. 
We further develop a formalized LED mapping methodology that combines full-band frequency sweep detector responses with synchronized LED activation to robustly assign resonator frequencies to physical locations. This includes a signal model for the detector response under illumination, a mapping algorithm for associating resonator shifts with LED positions, and a framework for quantifying mapping confidence and identifying failure modes. These elements enable a more systematic evaluation of mapping performance than has been previously reported.
Finally, we integrate the resulting resonator-position map into a corrective trimming pipeline, in which the measured frequency deviations are used to design lithographic modifications to the resonator capacitors. We present mapping results for one of the \qty{850}{GHz} arrays, including yield, frequency scatter, and assignment confidence, and discuss their implications for collision reduction and overall array performance.

This paper is organized as follows. Section 2 provides an overview of the \qty{850}{GHz} array and readout system. Section 3 describes the design and characterization of the LED mapper hardware. Section 4 details the LED mapping methodology and analysis framework. Section 5 presents the mapping results. Section 6 describes the integration of these results into the corrective trimming process, and Section 7 discusses the resulting impact on system performance. Conclusions are given in Section 9.

\section{CONTEXT}
\label{sec:context}

The \qty{850}{GHz} instrument module of Prime-Cam is designed to probe the highest-frequency band in the initial deployment of the FYST, extending the instrument's sensitivity into a part of the submillimeter regime where atmospheric transmission is restrictive but scientifically rich. The module is part of a broader system of multi-frequency cameras spanning approximately \qty{280}{GHz} to \qty{850}{GHz}, enabling simultaneous constraints on foregrounds, galaxy evolution, and early-universe structure formation \cite{vavagiakis2018, aravena2023}. The \qty{850}{GHz} (\qty{350}{\mu m}) focal plane is therefore required to achieve the highest submm detector density and efficient multiplexed readout while maintaining uniform performance across a large-format array to achieve unparalleled survey scan speeds \cite{aravena2023, gazda2025}. 

\subsection{Detector Array Architecture}

\begin{figure} [ht]
   \begin{center}
   \begin{tabular}{c} 
   \includegraphics[width=0.4\textwidth]{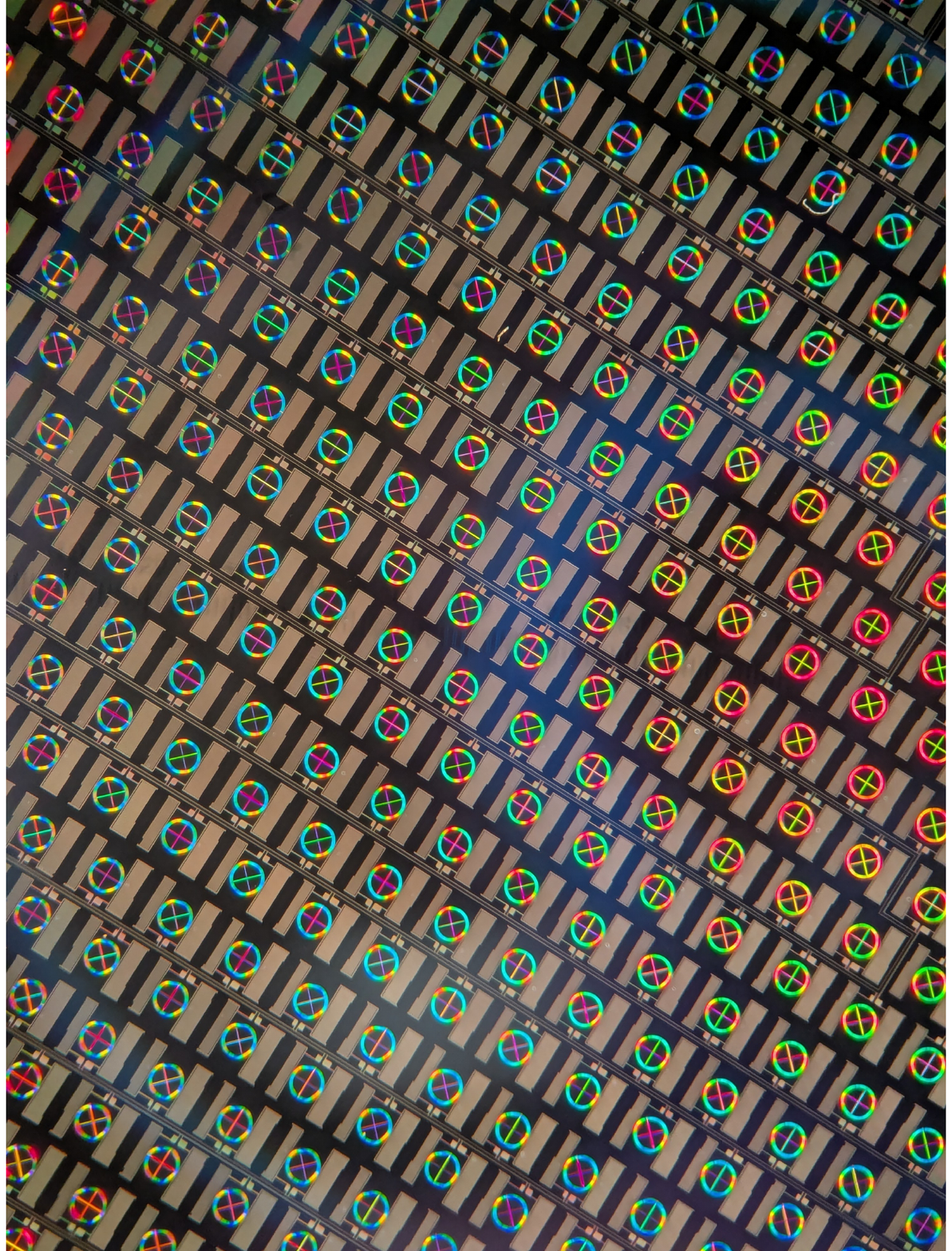}
   \includegraphics[width=0.4\textwidth]{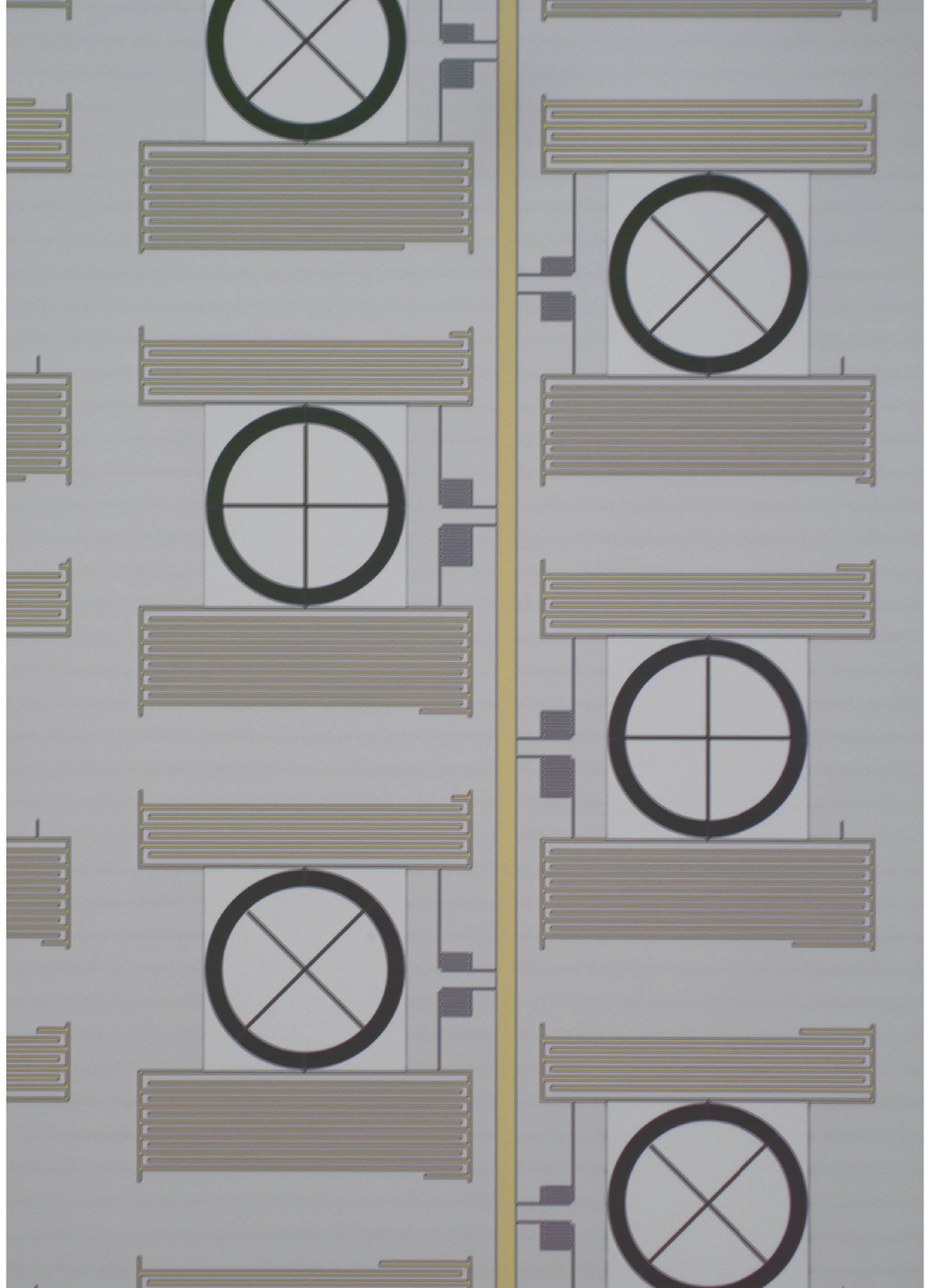}
   \end{tabular}
   \end{center}
   \caption[example] { 
       \label{fig:detectors} 
        (Left) A closeup view of the kinetic inductance detectors in this array. There are 12 different polarization alignments across the array. The pixel pitch is 1.4 mm.
        (Right) The `cross' in the center of each pixel are the orthogonal polarized inductive absorbers. The circular structure around the optically active inductor is dark inductance (only the `cross' is illuminated). The rectangular structure on either side of the circle are the interdigitated capacitors, with the coupling capacitors the much smaller rectangular structures near the feedline.
    }
\end{figure}

The detectors employed in the \qty{850}{GHz} module (see Fig.~\ref{fig:detectors}), fabricated by the Quantum Sensors Division at the National Institute for Standards and Technology (NIST), are feedhorn-coupled MKIDs, implemented as lithographically etched titanium-nitride (TiN) superconducting lumped-element resonator arrays whose resonant frequencies shift in response to absorbed optical power. Incident photons break Cooper pairs in the superconducting film, increasing the quasiparticle population and thereby modifying the kinetic inductance, which results in measurable shifts in both resonant frequency and quality factor \cite{zmuidzinas2012}. This mechanism enables direct detection of submm radiation with intrinsic frequency-domain multiplexing. 

Each detector is designed as a lumped-element resonator, in which the inductive element serves as the photon absorber and the capacitive element sets the resonant frequency. Arrays are fabricated such that each resonator has a unique designed frequency, allowing thousands of detectors to be read out on a single microwave feedline. Typical implementations attempt to space resonators in frequency within a limited bandwidth to avoid overlap, with the capacitor geometry varied across the array to achieve the desired frequency distribution. In practice, we implemented this distribution using logarithmic spacing defined by $f_n = f_0x^n$, where a geometric scaling factor $x$ slightly greater than unity (~1.001) provides a uniform fractional separation between adjacent channels.

The \qty{850}{GHz} module represents a significant scaling relative to earlier arrays. It has the highest pixel density and count of the Prime-Cam modules, and the most detectors per wafer of any submm receiver to date. The imaging plane is spanned by 3 hexagonal \qty{150}{mm} diameter silicon wafer TiN detector arrays, each with 6,348 dual-polarized pixels (12,696 detectors) split into 12 microwave networks for readout, yielding a total of 38,088 detectors. Each array is divided into 3 sub-sections of 4 networks, with detectors oriented to cover 12 different polarization angles \cite{huber2025}. 

The 1,058 detectors per network are spread over 2 readout octaves between \qty{0.2}{GHz} and \qty{1.2}{GHz}, placing stringent requirements on frequency placement accuracy and collision mitigation. This scaling directly motivates the need for robust mapping and post-fabrication correction to minimize collisions and maximize yield. The readout will utilize 1,024 resonators per network, leaving a buffer of approximately 3\% for maximum yield \cite{xie2024, sinclair2022, burgoyne2024, sinclair2023, sinclair2024}.

\subsection{Optical and Mechanical Integration}

The detector arrays are coupled to radiation through silicon-platelet feedhorn arrays, which defines the optical beam and couples incident power into the absorptive inductive elements of each KID. The feedhorn-coupled architecture provides well-controlled beam profiles and minimizes optical crosstalk between neighboring pixels, which is particularly important at high frequencies \cite{austermann2026}.

The focal plane assembly is integrated into a multi-stage cryogenic system within Prime-Cam. The instrument cryostat includes temperature stages at approximately 300 K, 80 K, 40 K, 4 K, 1 K, and a base temperature near 100 mK, achieved using a combination of pulse tube coolers and a dilution refrigerator \cite{vavagiakis2018}. The detector arrays are mounted at the lowest temperature stage to minimize thermal quasiparticle generation and achieve background-limited performance.

\subsection{Readout Architecture}

The readout of the MKID arrays is performed using microwave frequency-domain multiplexing \cite{sinclair2022}. A comb of probe tones, each corresponding to the designed resonant frequency of a detector, is generated at room temperature and transmitted through the feedline coupled to the array. Changes in the amplitude and phase of these tones, induced by shifts in resonator frequency and quality factor, are measured to recover detector signals.

Recent developments in readout electronics for Prime-Cam employ RF system-on-chip (RFSoC) platforms, which enable high-bandwidth, high-channel-count readout with reduced system complexity \cite{sinclair2023}. Current implementations are capable of reading out 1,024 detectors over 2 octaves up to 2 GHz, with multiple independent networks operating in parallel to scale to the full detector count \cite{xie2024, burgoyne2024}.

A key requirement of this architecture is that each resonator occupies a unique and well-separated frequency within the readout bandwidth. Fabrication-induced deviations from the designed frequencies introduce scatter that can lead to collisions between neighboring resonators, degrading readout performance and reducing usable detector yield. As detector counts increase, maintaining adequate frequency spacing becomes increasingly challenging, necessitating both precise characterization of the as-fabricated frequency distribution and corrective strategies to mitigate collisions.

\subsection{Motivation for LED Mapping in the 850 GHz Module}

The combination of large detector count, high multiplexing density, and fabrication variability makes it insufficient to rely solely on design-based frequency placement. Instead, it is necessary to empirically determine the mapping between measured resonant frequencies and physical detector locations on the array. This mapping enables identification of spatial patterns in frequency scatter, supports debugging of fabrication non-uniformities, and provides the essential input for post-fabrication correction through capacitor trimming in a re-etching process. 
For the \qty{850}{GHz} module, these requirements are amplified relative to lower-frequency arrays due to the increased detector density and tighter tolerances on frequency spacing. As a result, the LED mapping system described in this work is designed to provide high-fidelity, unambiguous association between resonators and physical pixels, forming a critical component of the overall detector calibration and optimization pipeline.

\section{LED ARRAY DESIGN}
\label{sec:led_array_design}

\subsection{Circuit Design and Schematic}

\begin{figure} [ht]
   \begin{center}
   \begin{tabular}{c}
   \includegraphics[width=0.8\textwidth]{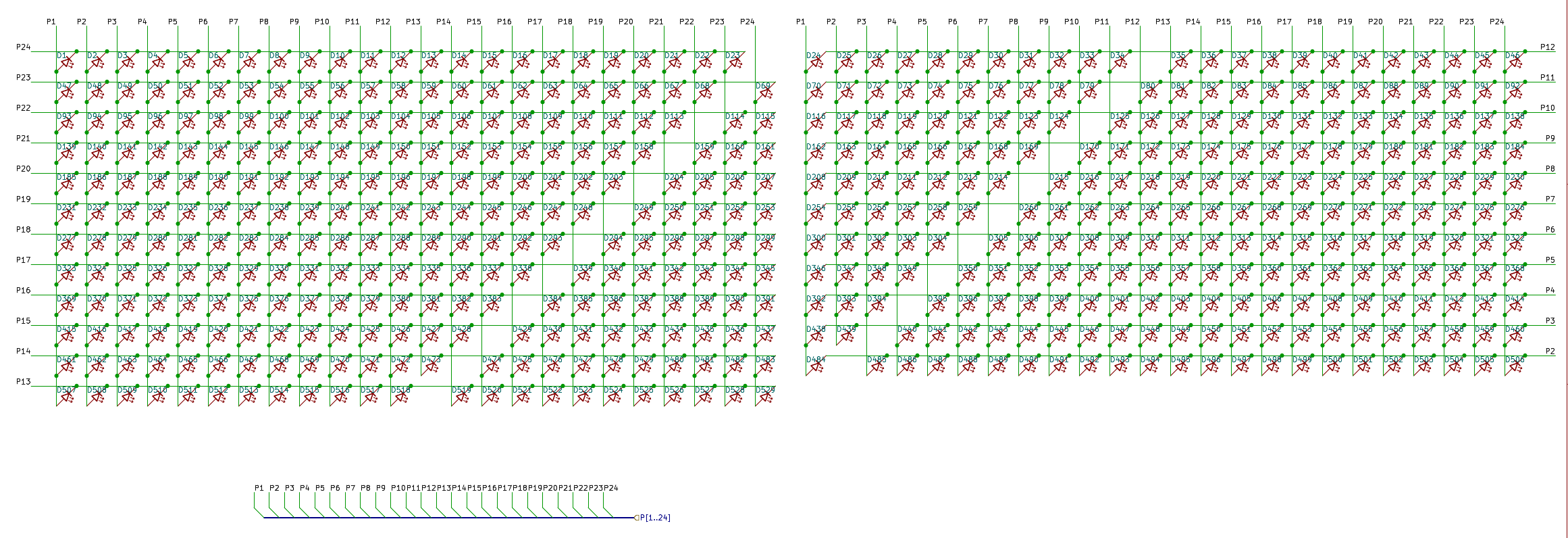}
   \end{tabular}
   \end{center}
   \caption[LED PCB] { 
       \label{fig:led_schematic} 
        KiCad schematic for a single 529-pixel network. A charlieplexed array can naturally be drawn as a square grid with n columns and n rows; however, this was split into two rectangles to match more closely to the physical layout of the network’s detectors.
    }
\end{figure} 

LEDs are coupled one-to-one with each dual-polarized pixel. To enable LED-resonator mapping, LEDs must be individually addressable within a network. A charlieplexed LED matrix leverages tri-state logic to enable individual addressability for large arrays with minimal input bus width. For a charlieplexed array with n wires, n(n-1) LEDs may be addressed individually, with two LEDs present at each wire junction, one of each polarity. To power an LED, the desired cathode and anode wires are set to logic high and logic low, respectively, and all other wires are set to high impedance mode. For a network size of 529 pixels, a set of 24 wires is sufficient to address all LEDs individually. The maximum number of individually addressable LEDs with 24 wires is 552, so some junctions are left open. Since the mapping of each network is performed independently, the same 24-wire bus may be used to address all networks simultaneously, where this effectively results in LEDs being powered in parallel, with one LED per network. Wiring LEDs in parallel can sometimes result in uneven power distribution due to nonlinearity, but LED luminance was observed to be well-distributed over multiple networks.

\subsection{LED Layout}

\begin{figure} [ht]
   \begin{center}
   \begin{tabular}{cc}
   \includegraphics[width=0.4\textwidth]{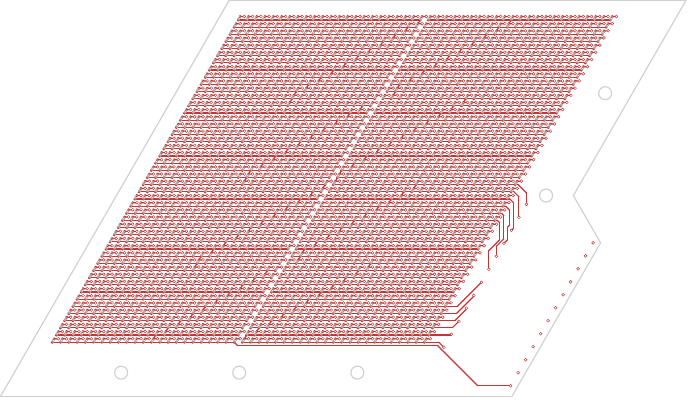} &
   \includegraphics[width=0.4\textwidth]{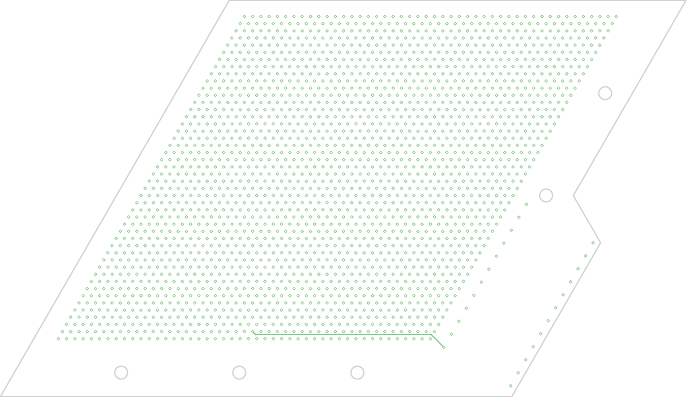} \\
   \includegraphics[width=0.4\textwidth]{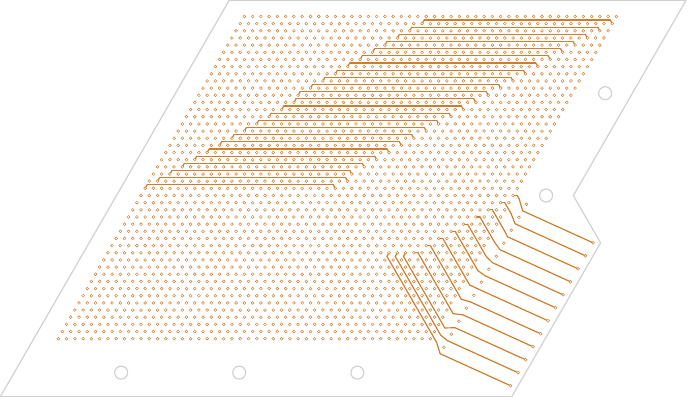} &
   \includegraphics[width=0.4\textwidth]{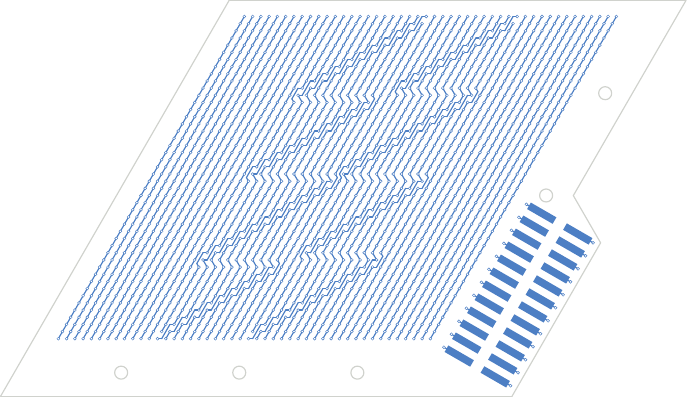}
   \end{tabular}
   \end{center}
   \caption[PCB traces] { 
       \label{fig:led_traces} 
        PCB traces shown in KiCAD for the four layers used in each LED mapping PCB. A) Front layer, on which the SMD LEDs are placed. Horizontal traces join cathodes for LEDs on the same network. B) Second layer, not used except to join one of the 24 wires to the connector. C) Third layer, with horizontal joining traces used to connect the vertical nets on the back layer, D) Back (fourth) layer, with vertical traces which connect anodes using through-hole vias and span all four networks in parallel.
    }
\end{figure} 

A primary constraint in the selection of Surface Mount Device (SMD) LEDs is the minimum center-to-center distance of 1.4 mm between pixels. The only commercially available standalone SMD LED size suitable for this application is the 0201 form factor. The package dimensions of the LEDs used are 0.65x0.35x0.4 mm (L/W/H) (Xinglight, 2026).

Red (630 nm) LEDs were selected due to their wide availability and low room-temperature forward voltage; because forward voltage increases at low temperatures, this low baseline ensures it does not exceed the 5 V maximum output of the Arduino.
While the positions of all LEDs are determined by the positions of the pixels on the detector array, the mapping from wire high/low pairs to physical locations within a given network is a free parameter, provided it is known. The final wiring implementation leverages the natural grid shape of the charlieplexed addressing system to minimize complexity.

The 24 addressing tracks each cross the entire array horizontally (rows) and vertically (columns). At each junction between two distinct wires, an LED can be addressed. Since each network is composed of 46 columns and 11.5 rows, the charlieplexed network is split into two rectangles: one 23x10 and one 23x11 section. The horizontal wires that connect the LED cathodes span the entire width of both rectangles, while the vertical wires that connect anodes are split in two. Additional joining traces on an inner layer are used to reconnect the vertical wires that are split across the two rectangles. Since the four networks on each PCB can share a common set of vertical traces, a single set of joining traces is sufficient to connect vertical wires for all networks. On the front layer, which primarily contains the horizontal traces, small traces are placed occasionally to connect through-hole vias to the horizontal traces, such that the correct horizontal and vertical wires are wired together for each of the 24 inputs.

\subsection{Overall circuit board}

\begin{figure} [ht]
   \begin{center}
   \begin{tabular}{c}
   \includegraphics[width=0.8\textwidth]{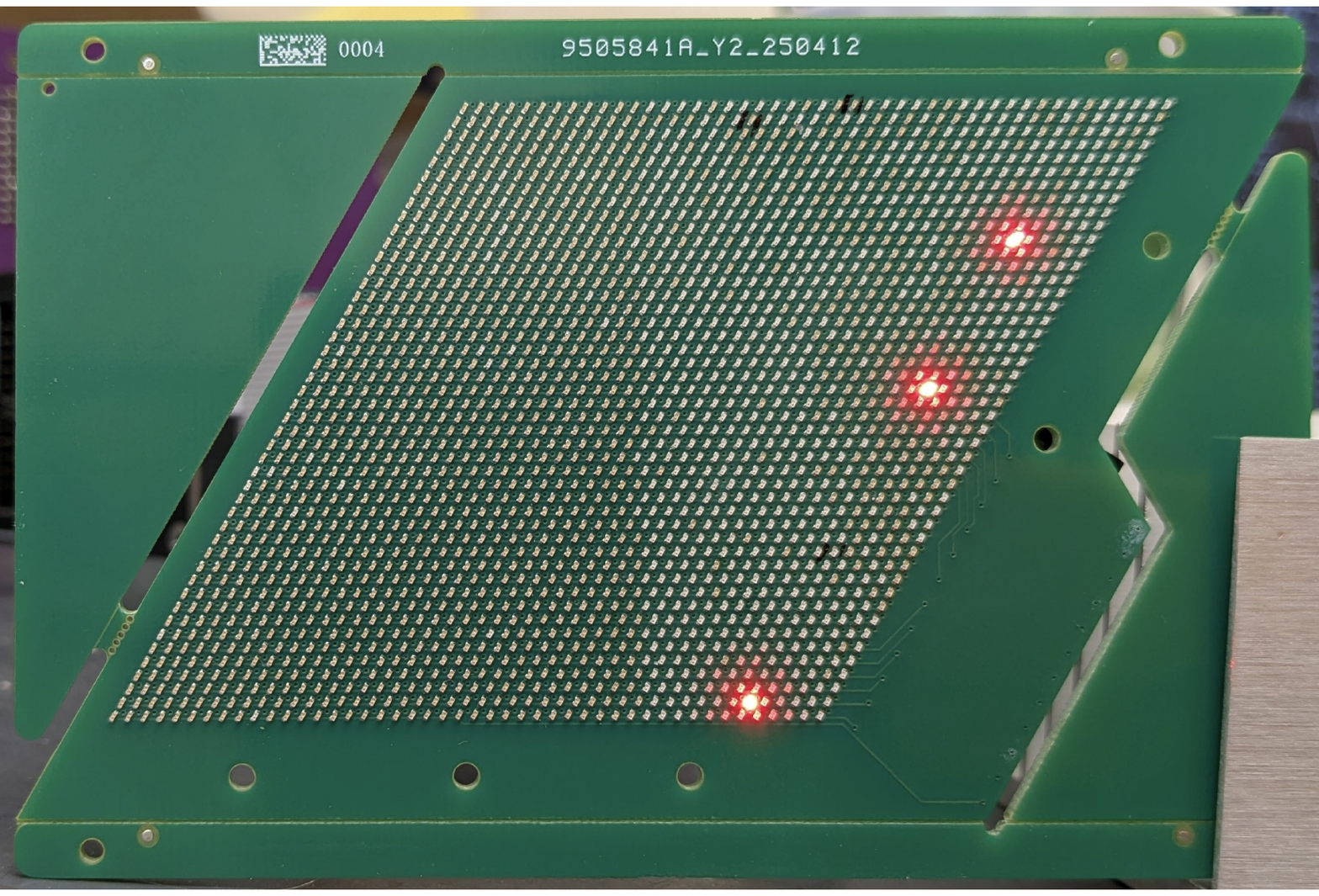}
   \end{tabular}
   \end{center}
   \caption[LED PCB] { 
       \label{fig:leds} 
        One of three LED PCBs needed to fully illuminate and array. The LEDs are Charlieplexed such that only a single LED can be powered in each network, with  four networks per LED PCB.
    }
\end{figure} 

The PCBs for each hexagonal array are subdivided into three rhombi, each one addressing four distinct detector networks. This subdivision reduces the risk of misalignment due to the difference in thermal contraction between the array and the LED mapping PCBs. Fewer, smaller boards also increase the proportion of LED mapping PCBs with no failed LEDs. One of three LED mapping PCBs are manufactured in a mirrored orientation, to align with the mirrored network on the array layout. A twenty-four pin connector is soldered manually onto each of the three PCBs, which is used to interface with the Arduino Mega.

\subsection{Collimator}

An aluminium collimator is mounted between the LED mapping PCBs and the MKID array to ‘collimate’ light and limit spillover to other pixels. 1 mm diameter holes are machined for each pixel, and MIC6 aluminum is used because of its high dimensional stability. A thickness 1/4’’ is used due to part availability. To reduce reflections and electrically insulate the collimator, it is anodized in black.
 
Since the aluminum collimator has a larger thermal expansion coefficient than the FR4 PCB, it is expected to shrink relative to the LED mapping PCB when cooled to sub-kelvin temperatures. A worst-case relative shift across the long diagonal of the rhombus was estimated at around 0.4 mm. The design accomodated the worst-case shift and the LED's were mated directly into the collimator holes to minimize optical cross-talk.

\section{MEASUREMENT PROCEDURE}
\label{sec:measurement_procedure}

\subsection{Instrumental Characterization}

\begin{figure} [ht]
   \begin{center}
   \begin{tabular}{c}
   \includegraphics[width=0.8\textwidth]{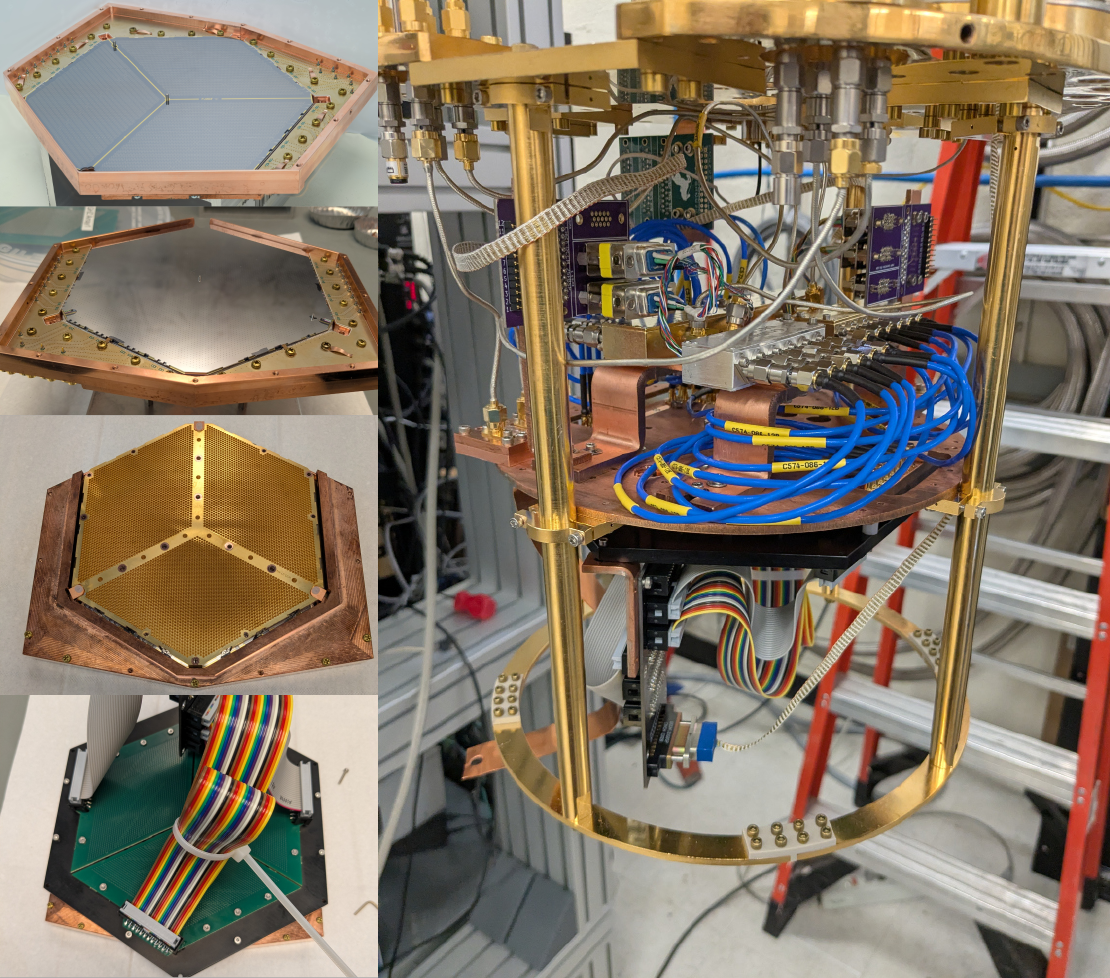}
   \end{tabular}
   \end{center}
   \caption[example] { 
       \label{fig:stack} 
        The optical stack for testing. On the left, from top to bottom, is the (mounted) detector array, the WIP, the feedhorns, and the LED setup. The LED setup includes 3 LED array PCBs mounted on a collimator array which reduces optical crosstalk going into the feedhorns.
    }
\end{figure}

This characterization work targets a single detector array (one of three in the focal plane), comprising 12 parallel networks with 1,058 dual-polarization frequency-multiplexed detectors per line (529 pixels). The testing optical stack assembly (see Fig.~\ref{fig:stack}) is made up of the LED electronics, LED and collimator arrays, feed horn array, waveguide interface plate, and detector array. The collimator assembly is machined from black-anodized 7075 aluminum, featuring a hole grid matched to the LED spacing. Aperture diameters (1.0 mm) are sized for 0201 LEDs with placement and drilling uncertainty and to fit within the pixel pitch (1.4 mm) while leaving enough material for rigidity (to minimize warping) and to maximize collimation and suppress optical crosstalk between adjacent feed horns. This stack is thermally isolated within the cryostat using stainless-steel spacers to maintain the thermal budget during mapping. LEDs are heat sunk at 1 K and thermally isolated via spacers to 100 mK.

Measurements are conducted in a cryogenic testbed at sub-Kelvin temperatures (approximately \qty{100}{mK}) to ensure stable superconducting behavior and mimic intended science operating conditions. The stack is mounted on the cold stage of a dilution refrigerator with careful thermal anchoring and magnetic shielding to suppress excess quasiparticle generation and frequency drift. The RF readout employs a coherent, frequency-swept multi-tone probe generated at room temperature, attenuated and thermalized along the input line before coupling to the detector feedline via impedance-matched coaxial wiring. The transmitted signal is amplified by a low-noise cryogenic amplifier at the \qty{4}{K} stage, followed by additional room-temperature gain and digitization. Complex transmission $S_{21}$ is acquired with a multi-tone bespoke vector network analyzer (built on an AMD Xilinx ZCU111), providing amplitude and phase across the resonator band. Readout power is set below the bifurcation threshold while maintaining sufficient SNR to resolve small LED-induced perturbations. Gain stability, cable delay, and standing-wave structure are calibrated to enable precise identification of resonance minima and frequency shifts.

The LED mapping sequence is driven by an Arduino Mega interfaced through a custom shield that routes logic pins to cryostat feedthroughs. Addressing uses a Charlieplexing architecture: a selected LED is driven by biasing its anode-cathode pair while all other pins are held in high impedance. To ensure reliable cryogenic operation and mitigate carrier freeze-out, the LEDs are cyclically stimulated during cooldown. The LEDs are in series with \qty{10}{k\Omega} resistors to set the optical power appropriate for these resonators and our testing needs, e.g. to achieve reasonable responses. Spatial identification of resonator-pixel pairs is obtained by comparing baseline sweeps with single-LED activations, mapping physical detector locations to their corresponding frequency-domain resonances.

\subsection{Data Acquisition and LED Modulation}

\begin{figure} [ht]
   \begin{center}
   \begin{tabular}{c}
   \includegraphics[width=0.8\textwidth]{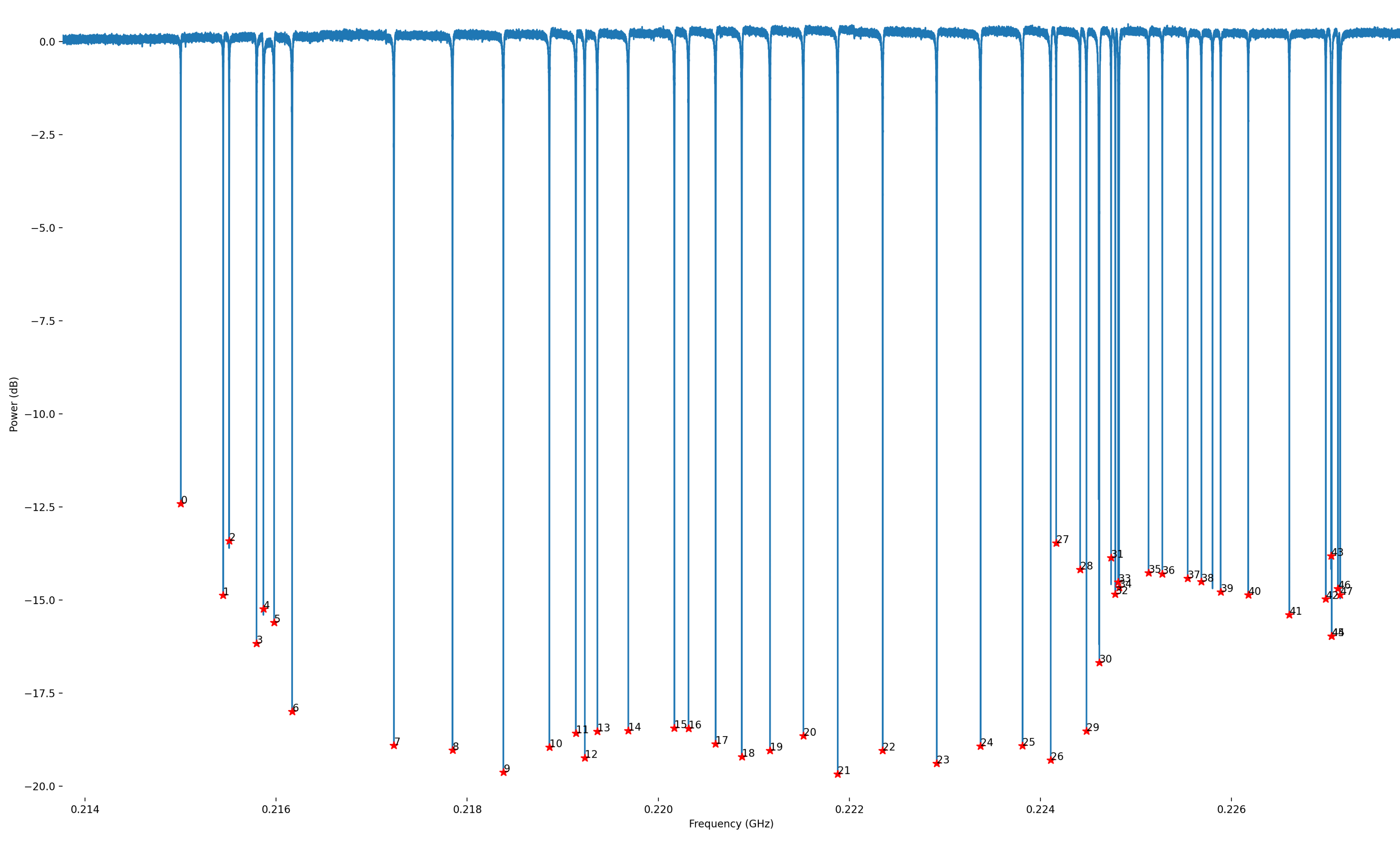}
   \end{tabular}
   \end{center}
   \caption[example] { 
       \label{fig:sweep_sample} 
        A sample of a frequency sweep of one of the networks (1) visualized in magnitude after fabrication but before trimming. The frequencies are designed to be evenly spaced but fabrication limitations result in some scatter.
    }
\end{figure}

Data acquisition is structured to ensure deterministic alignment between frequency-domain sweeps and LED states, with sampling density sufficient to resolve resonance perturbations. For each cycle, a baseline sweep $z_{\rm off}(f)$ is acquired with all LEDs off (see Fig.~\ref{fig:sweep_sample}), followed by a sequence $z_{\rm LED,i}(f)$ with individual LEDs activated in a serialized pattern. The frequency grid and per-point integration time are identical across all states, ensuring that inter-sweep differences reflect optical loading rather than instrumental drift. The frequency resolution oversamples the resonator linewidths, enabling robust localization of minima in $|z(f)|$ within windows around each candidate resonance. Systematics and random noise are reduced through post-processing, preserving relative depth and frequency shifts. The result is a consistent multi-state dataset suitable for differential analysis.

\subsection{Resonator Identification}

Resonator candidates are identified from the magnitude response using a thresholded, baseline-subtracted local-minimum detector tuned to narrow absorption features. The measured $|S_{21}|$ (in dB) is high-pass filtered by subtracting a low-pass, cosine-filtered estimate of the continuum, suppressing large-scale spectral structure and isolating localized dips. A fixed negative threshold is applied to the residual, and contiguous sub-threshold regions are segmented as candidate intervals. Within each interval, local minima are identified via a three-point comparison, requiring each candidate to be lower than its immediate neighbors. To limit over-detection in crowded regions, a minimum frequency spacing is enforced: pairs of minima separated by less than a specified threshold are treated as one, and only the deeper feature is retained. The procedure is embedded in an interactive visualization that allows in situ adjustment of the smoothing scale and detection threshold to obtain a robust initial set. Final curation removes spurious detections and recovers missed features. The surviving minima define resonator indices; fitting windows are constructed by partitioning the spectrum at midpoints between adjacent minima with configurable padding to ensure adequate coverage for subsequent model fitting.

\subsection{LED-Resonator Mapping}

LED-resonator association is performed by exploiting differential responses under controlled optical excitation and embedding the association problem in a low-dimensional metric space. For each candidate resonator, a local frequency window is extracted and the complex transmission is evaluated for both LED-off and LED-on states. Two summary statistics are computed per LED: (i) a relative index shift, defined as the offset between the nominal resonator index and the frequency sample at which the LED-on magnitude attains its minimum within the sliced window, and (ii) the corresponding minimum magnitude itself.

\begin{figure} [ht]
   \begin{center}
   \begin{tabular}{c}
   \includegraphics[width=0.8\textwidth]{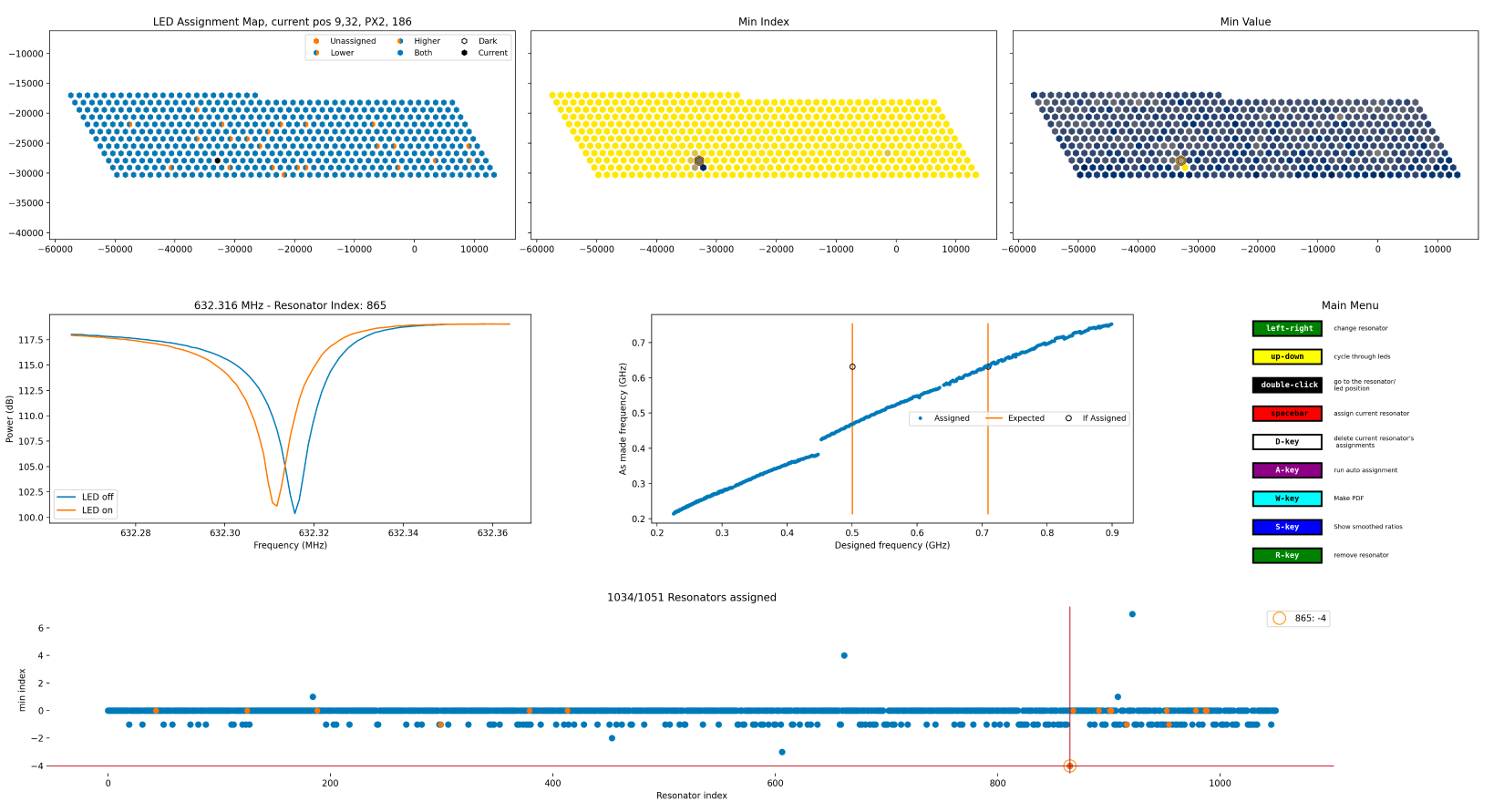}
   \end{tabular}
   \end{center}
   \caption[example] { 
       \label{fig:identify_interface} 
        The custom LED-resonator mapping interface. The top three plots show the LED (pixel) grid for a single network (1). Top left is the LED Assignment Map which displays which LEDs have already had resonator[s] assigned. Top middle is Min Index which shows which LEDs this resonator responded most strongly to. Top right is Min Value which illustrated the response of the selected resonator to the LEDs. The left plot on the center row shows the selected resonator in power as a function of frequency, with the blue line the baseline and the orange line the sweep when the selected LED was on. The right plot on the center row displays designed frequency against measured frequency for assigned resonators, demonstrating as expected pattern. The bottom plot shows which resonators have not yet been assigned (orange).
    }
\end{figure}

These quantities define per-resonator vectors over the set of LEDs, encoding which illumination condition produces the strongest perturbation and how localized that perturbation is in frequency. The resulting matrices are visualized as spatial heat maps over the physical pixel layout and as per-resonator traces. Candidate correspondences are identified via minima in the index-shift metric, with the magnitude metric serving as a consistency check against shallow or spurious responses. See Fig.~\ref{fig:identify_interface} for the user interface used in the manual part of the process.

The mapping procedure is specifically designed to resolve two common hardware-level ambiguities:
\textit{Spectral overlap}: Resonators that appear completely overlapped in the baseline $z_{\rm off}(f)$ sweep, often appearing as a single or distorted dip, are identified when the corresponding LED stimulates a frequency shift. This differential response pulls the resonators apart in the frequency domain, allowing them to be differentiated and indexed as distinct features.
\textit{LED failure and optical spillover}: In instances where a small fraction of LEDs are non-operational, the system leverages optical spillover. Because the illumination from a neighboring LED can still stimulate a marginal response in a resonator, dead LEDs can be identified in the Min Index and Min Value plots as clusters of LEDs stimulating a single resonator. In cases of ambiguity regarding which LED in a cluster has the primary stimulating effect, the relationship between designed and measured frequencies is used as a final arbiter. By plotting the measured resonance against the expected design pattern, the correct LED-resonator pair is identified by its fit within the global focal plane geometry.

An automated pass provisionally assigns matches by thresholding the index-shift metric and selecting the minimizing LED. Multi-resonator pixel sub-indices are inferred by proximity to design frequencies after correcting for a global frequency offset. All assignments are subsequently validated through conflict checks to prevent duplicate mappings. Finally, the data is refined through manual inspection using coordinated views including heat maps, frequency sweeps, and design-measurement planes to resolve remaining ambiguities and remove pathological resonators.

\section{MAPPING RESULTS}
\label{sec:mapping_results }

\begin{table}[ht]
    \centering
    \caption{
        LED mapping results for each network on the detector array. The 'Scatter' is defined as the normalized fractional offset between measured and designed resonance frequency. Dark collisions are determined as the fraction of resonators within 5 line widths of another resonator under unloaded conditions.
    }
    \begin{tabular}{ccccc}
    \hline
    Network & Designed [\#] & Measured [\#] & Scatter [\%] & Dark Collisions [\%] \\ \hline
    1  & 1052  & 1044  & 2.9 & 15.2 \\
    2  & 1056  & 1053  & 3.0 & 8.6 \\
    3  & 1056  & 1052  & 3.4 & 11.4 \\
    4  & 1056  & 1056  & 3.1 & 8.3 \\
    5  & 1056  & 1053  & 3.0 & 15.9 \\
    6  & 1056  & 1054  & 6.2 & 10.5 \\
    7  & 1056  & 1056  & 3.1 & 10.7 \\
    8  & 1056  & 1055  & 3.1 & 11.2 \\
    10 & 1056  & 1055  & 3.0 & 11.2 \\
    11 & 1056  & 1050  & 3.0 & 8.3 \\
    12 & 1052  & 1048  & 4.0 & 12.0 \\ \hline
    All & 11608 & 11576 & 3.4 & 11.1 \\ \hline
    \end{tabular}
\end{table}

\subsection{Mapping Yield}

The LED mapping procedure achieved an exceptionally high yield across the 850 GHz array. Out of 11,608 designed resonators across the 11 active networks, 11,576 candidates were successfully identified and mapped to physical pixels, representing a total yield of 99.7\%. The mean number of measured resonators per network was $1052.6 \pm 3.9$, closely matching the design mean of $1055.3 \pm 1.5$ (10 of the networks were designed for 1056 resonators, and 2 were designed for 1052 resonators).

It should be noted that Network 9 is excluded from these statistics; a significant surface scratch on the feedline resulted in severe signal attenuation, rendering the network unusable for standard operations. While mapping of Network 9 was eventually achieved through targeted high-power sweeps and manual verification, it is omitted here to provide a representative view of nominal detector performance.

\subsection{Frequency Scatter}

\begin{figure} [ht]
   \begin{center}
   \begin{tabular}{c}
   \includegraphics[width=1\textwidth]{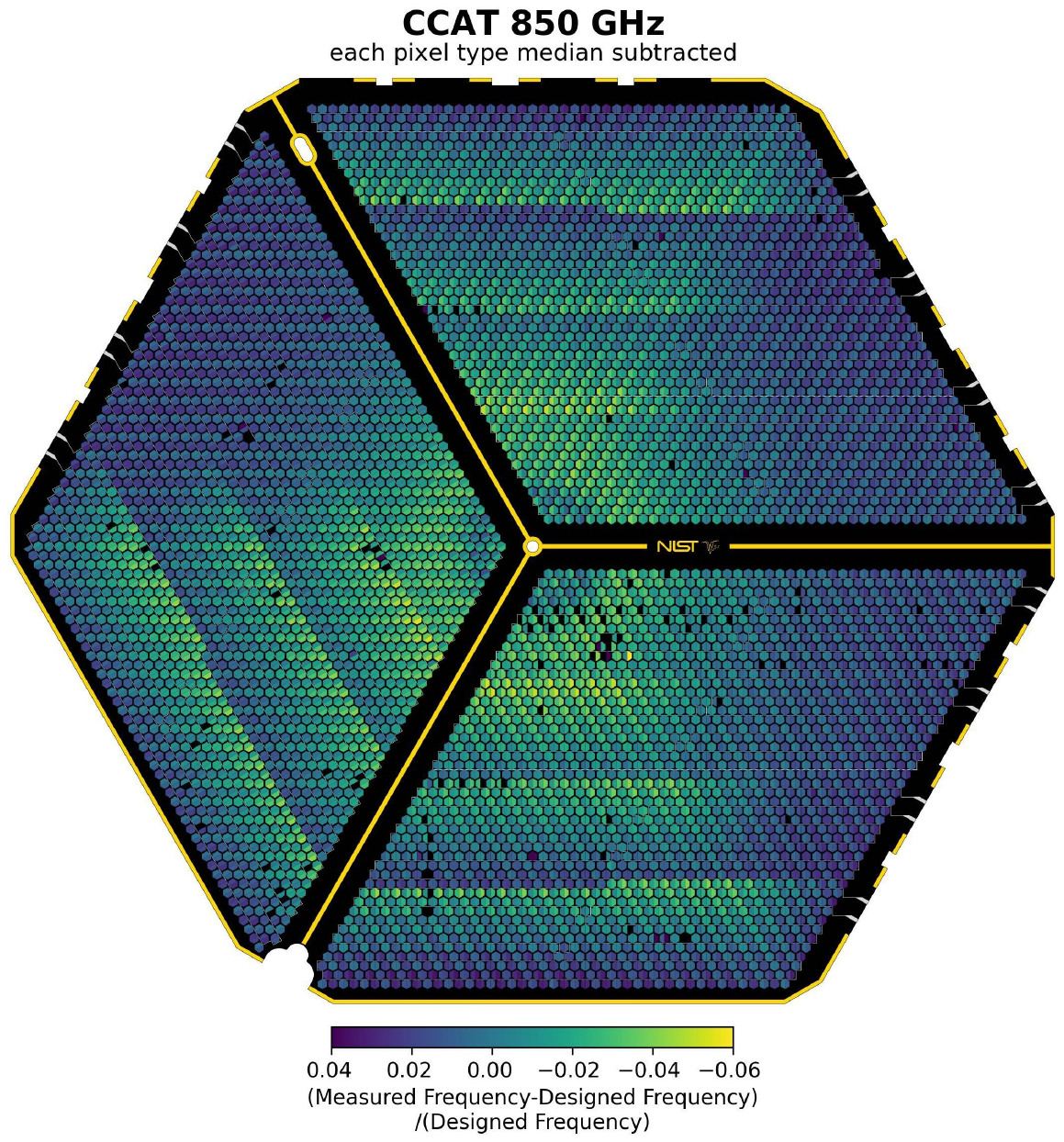}
   \end{tabular}
   \end{center}
   \caption[example] { 
       \label{fig:mapped_detectors} 
        Spatial distribution of fractional frequency offsets across the 850 GHz array. The color scale indicates the normalized deviation between measured and designed resonance frequencies.
    }
\end{figure}

The frequency scatter, defined as the normalized fractional offset between measured and designed resonance, can be seen in Fig.~\ref{fig:mapped_detectors}. Frequency scatter has a mean value of $3.43 \pm 0.92\%$ across the full focal plane. This tight distribution indicates high fabrication fidelity, though localized variations are present. Specifically, Network 6 exhibited a higher scatter of 6.2\%, contributing to the distribution's tail.

Spatial mapping of these fractional frequency offsets reveals coherent, large-scale geometric gradients across the focal plane tiles. Two distinct macroscopic patterns emerge: an intra-network gradient and a systematic diagonal ``striping''. The intra-network gradients manifest as twelve distinct rhombic profiles, physically aligning with the network layouts. This variation is driven by frequency-dependent geometric non-uniformities in the interdigital capacitors, where the finger and gap dimensions were deliberately varied during design optimization to balance performance across the band. Conversely, the diagonal striping pattern is a fabrication artifact, likely resulting from subtle geometric inductance disparities between shorted and unshorted detectors. These spatial structures indicate that while localized fabrication fidelity remains high, macroscopic variations in both lithographic reproduction and geometric design choices systematically modulate the pixel kinetic inductance, inducing predictable shifts from the target design resonances.

\subsection{Collisions}

\begin{figure} [ht]
   \begin{center}
   \begin{tabular}{c}
   \includegraphics[width=0.8\textwidth]{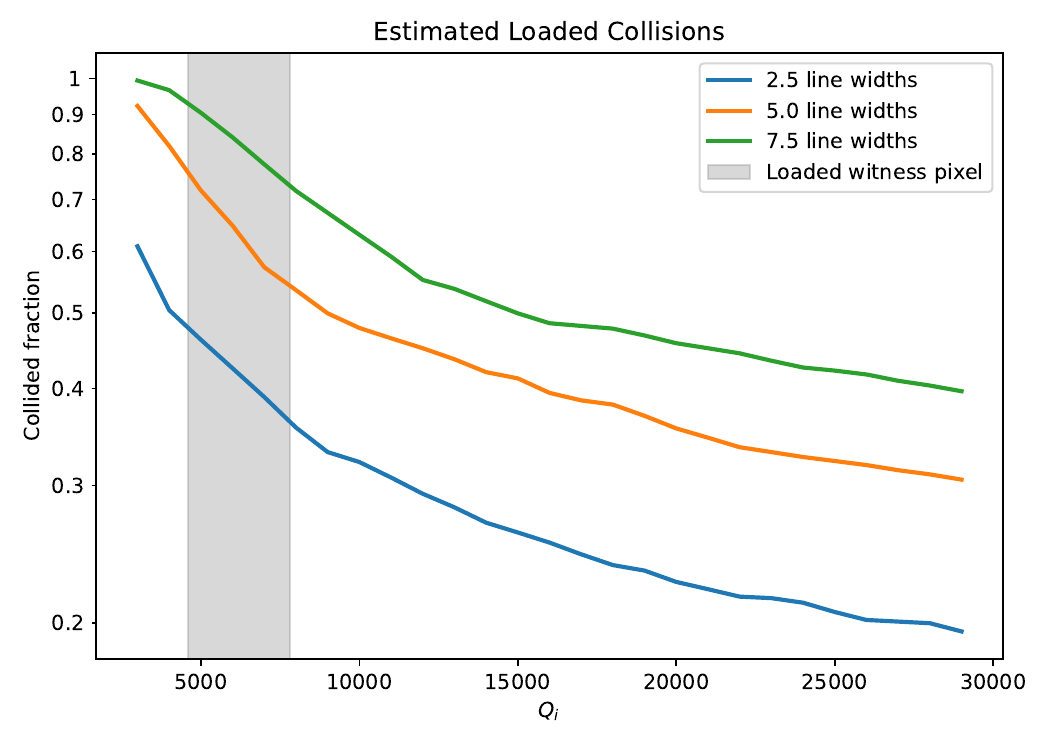}
   \end{tabular}
   \end{center}
   \caption[example] { 
       \label{fig:collisions} 
       Estimated detector frequency collision fraction under expected loading. The colored lines define a collision as two detectors being within 2.5, 5.0, or 7.5 line widths of each other. The grey band represents the internal quality factor of the detectors under expected loading, determined from characterization of a witness pixel.
    }
\end{figure}

Spectral collisions are a primary challenge in high-density frequency-domain multiplexing. For this analysis, a detector is defined as collided if another resonance exists within five linewidths at a lower frequency in the baseline $z_{/r off}(f)$ sweep. We use five linewidths to ensure minimal crosstalk; as the gap reduces, crosstalk will increase. Across the 850 GHz array, we observed a mean dark (no loading) collision rate of $11.1 \pm 2.6\%$. Under expected loading conditions the internal quality factor will decrease, causing an increase in the number of collisions. From characterization of a witness pixel we estimate the collided fraction on this array under expected loading conditions will be in the neighbourhood of 60\% (see Fig.~\ref{fig:collisions}), before corrective trimming, due to a collusion of sub-optimal $Q_c$ and $Q_i$ in this first array. For the next arrays, we expect the corrective trimming process to reduce the loaded collision rate to less than 10\%.

Despite these spectral overlaps and the presence of non-operational ``dead'' LEDs, the mapping algorithm robustly resolved LED-resonator associations by leveraging optical spillover from neighboring pixels. In cases where an LED failed to trigger, the surrounding LEDs often provided enough incidental illumination to stimulate a measurable frequency shift. By identifying the specific cluster of LEDs that perturbed a single resonator, the correct assignment was determined by selecting the LED that minimized the deviation from the expected geometric design pattern.

This mapping process was significantly bolstered by the high fabrication uniformity of the Titanium Nitride (TiN) film. The variability between designed and measured resonant frequencies in TiN has been substantially lower than in Aluminum (Al) in our testing, allowing for much tighter constraints when fitting resonators to their predicted frequency-spatial coordinates.

Finally, a small fraction of resonators, typically those displaying unstable baselines or insufficient SNR for robust LED stimulation, were flagged as pathological and removed during the manual curation phase to ensure the integrity of the final detector catalog (mean of 3).

\subsection{Quality Factors}

The coupling quality factor $Q_c$ exhibits a distinct alternating-row pattern across the array. This spatial variation is driven by the routing architecture, where the readout feedline is positioned between every second pixel row, requiring adjacent detectors to couple to the line from opposite sides. Because the interdigital capacitor fingers are oriented along a top-and-bottom topology, and only one side was trimmed during the primary lithographic etching process, a systematic asymmetry in total finger length was introduced between the upper and lower halves of the structure. Consequently, pixels coupling to the feedline from the top experience a measurably different capacitive coupling, and thus a shifted $Q_c$, compared to those coupling from the bottom.



\section{CONCLUSION}
\label{sec:conclusion}

We have designed, implemented, and characterized a cryogenic LED mapping system for the 850 GHz instrument module of Prime-Cam on the Fred Young Submillimeter Telescope (FYST). The Charlieplexed hardware architecture successfully addressed the high-density detector channels with minimal input wiring, while the black-anodized aluminum collimator effectively suppressed optical crosstalk. Cryogenic evaluation at 100 mK demonstrated an exceptionally high mapping yield of 99.7\%, successfully pairing 11,576 resonators across 11 active networks.

The analysis revealed a mean fractional frequency scatter of 3.43 ± 0.92\%, exposing macroscopic geometric gradients and lithographic striping artifacts across the focal plane. Additionally, the spatial distribution of the coupling quality factor ($Q_c$) tracked systematic asymmetries driven by feedline routing topology. The differential mapping algorithm proved highly robust, successfully resolving overlapping resonances and mitigating localized LED failures by leveraging optical spillover from adjacent pixels. The resulting high-fidelity spatial-frequency coordinate maps provide the essential empirical baseline required for the subsequent post-fabrication lithographic capacitor trimming pipeline, ensuring the mitigation of frequency collisions and maximizing the operational detector yield for submillimeter observations.

\acknowledgments 
 
The CCAT project, FYST and Prime-Cam instrument have been supported by generous contributions from the Fred M. Young, Jr. Charitable Trust, Cornell University, and the Canada Foundation for Innovation and the Provinces of Ontario, Alberta, and British Columbia. The construction of the FYST telescope was supported by the Gro{\ss}ger{\"a}te-Programm of the German Science Foundation (Deutsche Forschungsgemeinschaft, DFG) under grant INST 216/733-1 FUGG, as well as funding from Universit{\"a}t zu K{\"o}ln, Universit{\"a}t Bonn, the Max Planck Institut f{\"u}r Astrophysik, Garching, and Duke University.

The construction of EoR-Spec is supported by NSF grant AST-2009767. The construction of the 350 GHz instrument module for Prime-Cam was supported by NSF grant AST-2117631.
funds.

\bibliography{report} 
\bibliographystyle{spiebib} 

\end{document}